# COMPACT WEIGHTED CLASS ASSOCIATION RULE MINING USING INFORMATION GAIN

S.P.Syed Ibrahim[1] and K.R.Chandran[2]

[1]Assistant Professor, Department of Computer Science and Engineering,
PSG College of Technology, Coimbatore, India
`sps_phd@yahoo.co.in`

[2]Professor of Information Technology and Head, Department of Computer and Information Sciences, PSG College of Technology, Coimbatore, India
`chandran_k_r@yahoo.co.in`

## ABSTRACT

*Weighted association rule mining reflects semantic significance of item by considering its weight. Classification constructs the classifier and predicts the new data instance. This paper proposes compact weighted class association rule mining method, which applies weighted association rule mining in the classification and constructs an efficient weighted associative classifier. This proposed associative classification algorithm chooses one non class informative attribute from dataset and all the weighted class association rules are generated based on that attribute. The weight of the item is considered as one of the parameter in generating the weighted class association rules. This proposed algorithm calculates the weight using the HITS model. Experimental results show that the proposed system generates less number of high quality rules which improves the classification accuracy.*

## 1. INTRODUCTION

Data mining principally deals with extracting knowledge from data, In the world where data is all around us, the need of the hour is to extract knowledge or interesting information, which is hidden in the available data. Association rule mining is concerned with extracting a set of highly correlated features shared among a large number of records in a given database. It uses unsupervised learning where no class attribute is involved in finding the association rule. Although classical association rule mining algorithm reflects the statistical relationship between items, it does not reflect the semantic significance of the items [10]. To meet the user objective and business value, various weighted association rule mining methods [17] [15] [16] were proposed based on the weightage to items.

On the other hand, classification uses supervised learning where class attribute is involved in the construction of the classifier. Both, weighted association rule mining and classification are significant and efficient data mining techniques. So integration of these two data mining techniques may provide efficient associative classifier [18].

In [18] syed et al., have proposed weighted associative classification based on class based association (CBA) [13]. Weighted associative classification (WAC) algorithm pre-assigns the weight for each item randomly. WAC algorithm generates huge number of rules, so it suffers with high computation cost.

To improve the performance, recently the authors proposed compact weighted associative classification (CWAC) [19]. The CWAC algorithm is completely varies from WAC. In WAC, Apriori association rule mining algorithm is directly applied to find the class association rules, whereas CWAC algorithm chooses one non class attribute randomly from dataset and all the items are generated only based on that attribute. In this way CWAC algorithm reduces the number of itemset generation and it assigns the weight for each item using HITS model [11]. CWAC algorithm calculates the weighted support and weighted confidence for each item and determines whether the item was frequent or not. This paper aiming to improve the performance of the CWAC algorithm and evaluates the effect of weighted associative classification method in the various bench mark datasets.

The rest of the paper is organized as follows: Section 2 gives an insight about the past work in this field and section 3 explains the attribute selection strategy, rule generation and rule evaluation strategies. Section 4 presents the experimental results and observations followed by the conclusion.

## 2. RELATED WORK

### 2.1 Association Rule Mining

Association Rule Mining (ARM) [1], has become one of the important data mining tasks. ARM is an unsupervised data mining technique, which works on variable length data, and it produces clear and understandable rules. The basic task of association rule mining is to determine the correlation between items belonging to a transactional database. In general, every association rule must satisfy two user specified constraints, one is support and the other is confidence. The support of a rule $X \rightarrow Y$ (X and Y are items) is defined as the fraction of transactions that contain X and Y, while the confidence is defined as the ratio support(X and Y)/support(X). So, the target is to find all association rules that satisfy user specified minimum support and confidence values.

### 2.2. Weighted Association Rule Mining

Classical ARM framework assumes that all items have the same significance or importance i.e. their weight within a transaction or record is the same (weight=1 per item) which is not always the case. In the supermarket context, some items like jewellery, designer clothes, etc., are of much significant in terms of revenue or profit by the store. Hence weight can be used as a parameter to generate association rule mining called as weighted association rule mining [17, 16]. The weighted association rules are generated based on user specified minimum weighted support and minimum weighted confidence thresholds. The use of weighted support and weighted confidence leads to useful mechanisms to prioritize the rule according to their importance, instead of their support and confidence alone.

### 2.3. Classification

Construct the classifier based on training dataset and predicts the class object for new dataset. Classification uses supervised learning where class attribute is involved in constructing a classifier.

### 2.4. Associative Classification

Associative classification was first introduced by Liu et al[13] which focus on integrating two known data mining tasks, association rule discovery and classification. The integration done is focused on a special subset of association rules whose right hand side is restricted to the class attribute; for example, consider a rule R: $X \rightarrow Y$, Y must be a class label. Associative classification generally involves two stages. In the first stage, it adopts the association rule generation methods like Apriori candidate generation [2], or FP growth [8] algorithms to generate class association rules.

For example CBA [13] method employs Apriori candidate generation [2] and other associative methods such as CPAR [24], CMAR [12] while Lazy associative classification [3, 4] methods adopts FP growth algorithm [8] for rule generation. The rule generation step generates huge number of rules. Experimental results reported by Baralis et al., [4] has shown that the CBA method which follows Apriori association rule mining algorithm generates more than 80,000 rules for some datasets that leads to memory exceptions and other severe problems, such as overfitting etc.,

If all the generated rules are used in the classifier then the accuracy of the classifier would be high but the process of classification will be slow and time-consuming. So in the next stage, generated rules are ranked based on several parameters and interestingness measures such as confidence, support, lexicographical order of items etc. Then only the high-ranking rules are chosen to build a classifier and the rest are pruned.

Evolutionary based associative classification method [22] was proposed recently. This approach takes subset of rules randomly to construct the classifier. Richness of the ruleset was improved over the generation.

Syed et al., [21] proposed Lazy learning associative classification where it classifies the new data sample without constructing the classifier but this lazy approach results in high CPU utilization time and cost.

Chen et al., [7] and Zhang et al., [25] proposed a new approach based on information gain where more informative attribute are chosen for rule generation. An informative attribute centred rule generation produces a compact ruleset.

Syed et al.,[20] proposed compact weighted associative classification based on information gain, where class association rule generation algorithm chooses information gain non class attribute from dataset and all the items are generated only based on that attribute. Thus this algorithm reduces number of itemset generation. Finally the algorithm calculates the weighted support and weighted confidence for each item and determines whether the item is frequent or not.

In [19] the author's proposed genetic network based associative classification method, which generates sufficient number of rules to construct the classifier. Here information gain attribute is used to construct the compact genetic network.

## 2.5. Weighted associative classification

Syed et al., [18] have proposed weighted associative classification (WAC), which integrates weighted association rule mining and classification to construct the efficient weighted associative classifier. Weighted associative classifier extracts special subset of association rules called weighted class association rules (WCARs). Weighted association rule mining uses weight as one parameter but here weights for each item item are assigned randomly but it is very difficult to assign weights to each item.

## 2.6. HITS Model

Kleinberg [11] used HITS algorithm in bipartite graph and weights are derived from the internal structure of the database. Sun et al., [17] uses this HITS model to derive weights for each item in the dataset and derived the weighted association rules. This proposed method uses HITS model to derive the weight for each item. Then these weights are used to compute the Weighted Class Association Rules.

## 3. PROPOSED SYSTEM

### 3.1. Problem Definition

Let database D is a set of instances where each instance is represented by < a1, a2 …am , C>, where a1, a2 …am, are attributes and C are class value each has weights. A common rule is defined as x → c, where x is a set of non class attributes and c is class label. The quality measurement factor of a rule is weighted support and weighted confidence. Rule items that satisfy minimum weighted support and weighted confidence are called frequent weighted rules, while the rest are called infrequent weighted rules. Here the task is to generate the Weighted Class Association Rules (WCARs) that satisfies both minimum weighted support and minimum weighted confidence constraints. Then these WCARs are used to construct a classifier based on Confidence, Support, and size-of-the rule Antecedent.

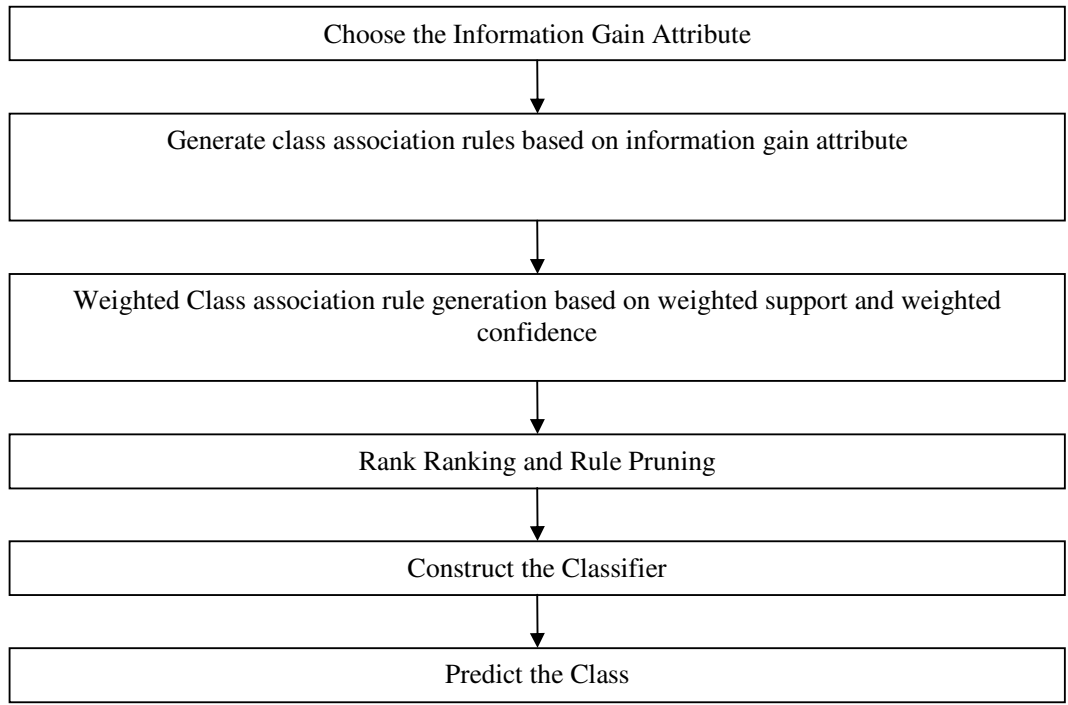

Figure 1. The proposed System

### 3.2 Attribute Selection based on Information Gain

Generally Apriori based rule generation algorithm generates 2k rules for the dataset with K items. To reduce the rule generation Information gain attribute is used. Information gain is a measure that is used in information theory to quantify the 'information content' of messages [9]. In ID3 decision tree algorithm [14] information gain is used to choose the best split attribute.

Information gain measure is used to identify the best split attribute in decision tree classifier. In this paper, it could be used to generate the class association rules. In the process of generating the class association rules, instead of considering all the attributes, information gain measure will be used to select the best splitting attribute. In this way, the proposed work generates the limited number of good quality rules.

Suppose an attribute A has n distinct values that partition the dataset D into subsets $T_1, T_2, ..., T_n$. For a dataset, $freq(C_k, D) / |D|$ represents the probability than an tuple in D belongs to class $C_k$. Then info(D) is defined as follows to measure the average amount of information needed to identify the class of a transaction in D:

$$\text{Info}(D) = -\sum_{k=1}^{g} \frac{freq(C_k, D)}{|D|} X \log_2 \frac{(freq(C_k, D))}{|D|} \quad (1)$$

Where |D| is the number of transactions in database D and g is the number of classes. After the dataset D is partitioned into n values of attribute A, the expected information requirement could be defined as:

$$\text{Info}_A(D) = \sum \frac{|D_i|}{|D|} X \text{ info}(D_i) \quad (2)$$

The information gained by partitioning D according to attribute A is defined as:

Gain (A) = info (D) – info A (D)  (3)

The best split attribute is the one that maximized the information gain in the data set D. These best attributes is used to generate the subset.

## 3.3 Subset Generation

After identifying the information gained attribute, the subsets are generated only based on information gain attribute. Let us consider Table I, contains 14 transaction, and 2 class values. Among the four attributes, attribute 'CD4 Cell Count' is selected as the informative attribute as it has the maximum IG value.

TABLE 1 Sample Dataset

| CD4 Cell Count | Sweating at Night | Tuberculosis (TB) | Temperature | AIDS |
|---|---|---|---|---|
| >500 | High | no | Normal | No |
| >500 | High | no | High | No |
| <200 | High | no | Normal | Yes |
| 200 .. 500 | Medium | no | Normal | Yes |
| 200 .. 500 | Nil | yes | Normal | Yes |
| 200 .. 500 | Nil | yes | High | No |
| <200 | Nil | yes | High | Yes |
| 200 .. 500 | Medium | no | High | No |
| >500 | Medium | no | Normal | No |
| >500 | Nil | yes | Normal | Yes |
| 200 .. 500 | Medium | yes | Normal | Yes |
| >500 | Medium | yes | High | Yes |
| <200 | Medium | no | High | Yes |
| <200 | High | yes | Normal | Yes |

So for the above dataset the following subset can be created for the scanning of the first transaction. Generated rules using information gain based associative classification are

{CD4 Cell Count >500} → AIDS =No}
{CD4 Cell Count > 500, Sweating at Night = High → AIDS = No}
{CD4 Cell Count > 500, Sweating at Night = High, Tuberculosis (TB) = No → AIDS= No}
{CD4 Cell Count > 500, Sweating at Night = High, Tuberculosis (TB) = No,
 Temperature == Normal → AIDS = No}

The other itemsets that are commonly generated by the Apriori based association rule mining procedure are eliminated. The excluded rules are

{Sweating at Night == High → AIDS = No }
{Tuberculosis == No → AIDS = No }
{Temperature == Normal → AIDS = No }
{Sweating at Night == High, Tuberculosis == No → AIDS = No }
{Sweating at Night == High, Temperature == Normal → AIDS = No }
{Tuberculosis == No,Temperature == Normal → AIDS = No }
{CD4 Cell Count > 500, Tuberculosis == No → AIDS = No }
{CD4 Cell Count >500, Temperature == Normal → AIDS = No }
{CD4 Cell Count > 500 , Sweating at Night == High, Temperature == Normal → AIDS = No }
{CD4 Cell Count >500, Tuberculosis == No, Temperature == Normal}
{Sweating at Night == High, Tuberculosis == No, Temperature == Normal}
This clearly shows this algorithm generates minimal number of rules.

## 3.4 Weighted class association rule generation

Weighted Associative classifier construction is of two steps. In the first step, all the weighted class association rules are generated based on weighted association rule mining technique. To find the weighted association rule, weight is the important factor. HITS model is applied to the dataset as in [17]. From this we can derive the weighted for each item.

Here weighted support and weighted confidence measures are used to evaluate the rule.

Definition 1: The w-support of a rule X→C is defined as

$$WSup(X \to C) = \sum \frac{Hub\ Weight(X \to C)}{Total\ Hub\ Weight} \qquad (4)$$

Definition 2: The w-confidence of a rule X→C is defined as

$$WConf(X \to C) = \frac{WConf(X \to C)}{WConf(X)} \qquad (5)$$

Class association rule is said to significant if its weighted support is larger than the minimum weighted support. Then the rules are ranked and the rules that satisfy certain threshold conditions are used to construct the classifier. After rule ranking, only the high-ranking rules are chosen to build a classifier and the rest are pruned.

### 3.5 Rule ranking and rule pruning

The minimum weighted support and minimum weighted confidence are user defined threshold values. The itemset that has weighted support and weighted confidence above the threshold value are called as frequent itemset and others are called as infrequent itemset which are pruned during rule generation process. If all the frequent rules are used in the classifier then the accuracy of the classifier would be high but the process of classification will be slow and time-consuming. So rule ranking and rule pruning techniques are proposed to choose an optimal rule set. To apply rule pruning, the generated rules are ranked based on several parameters and interestingness measures such as weighted confidence, weighted support, lexicographical order of items etc. Initially the rules are arranged based on their weighted confidence value. If two rules have the same value for the weighted confidence measure then the rules are sorted based on their weighted support. If both weighted confident and weighted support values are same for two rules then the sorting is done based on their rule length. Even after considering weighted confidence, weighted support, and cardinality measures, if there exists some rules with the same values for all three measures then the rules are sorted based on its lexicographic order as in Lazy pruning [4] method.

### 3.6 Conflict Rule Reduction

Even after applying rule pruning strategies their may exist conflicting rules and redundant rules.
If there exists two rules r1 and r2 where
    r1:X→Ci and r2:X→Cj
then r1 and r2 are said to be conflicting rules.

Let us consider a ruleset where, there exist two rules like
    r1:X→Ci and r2:X ^ Y →Ci
then r1 and r2 are said to be redundant rule as they are semantically meaningless from a classification viewpoint. In the gain based approach [7][25] confidence threshold was set greater than 50%. This further enhances the ruleset as conflicting rules and redundant rules are avoided. The remaining rules are used in the classifier.

### 3.7 Compact Weighted Associative Classification Algorithm

This section explains Compact weighted class association rule generation algorithm.

INPUT: DATA SET
OUTPUT: CLASS

I. Choose a non class informative attributet.
II. Generate one item- class association rules based on the selected attribute.
III. Calculate weighted support.
IV. If weighted support of item is greater than minimum weighted support then generate two itemset and so forth
V. After generating all the itemset calculate weighted confidence for all itemset

Classifier Algorithm

I. Rank the rules based on Weighted Confidence, Weighted Support and Size of the rule antecedent.
II. Classify the test dataset using these ruleset and obtain the classifier accuracy.

## 4. Experimental results

The proposed system was tested using benchmark datasets from the University of California at Irvine Repository (UCI Repository) [6]. The datasets were preprocessed to convert to a general format. A brief description about the datasets is presented in Table 3. The experiments were carried out on a PC with Intel Core 2 Duo CPU with a clock rate of 1.60 Ghz and 2 GB of main memory. Holdout approach [9] was used to randomly choose the training and testing dataset from the dataset. The training dataset is used to construct a model for classification. After constructing the classifier, the test dataset is used to estimate the classifier performance.

### 4.1 Accuracy Computation

Accuracy measures the ability of the classifier to correctly classify unlabeled data. It is the ratio of the number of correctly classified data over the total number of given transactions in the test dataset.

$$Accuracy = \frac{Number\ of\ objects\ correctly\ classified}{Total\ number\ of\ objects\ in\ the\ testset} \quad (6)$$

The performance of the proposed compact weighted class association rule mining method was evaluated by comparing it with the traditional associative classification algorithm (CBA) [13] and the existing compact class association rule mining (GARC) [7]. Table 3 gives the dataset description for various datasets.

TABLE 3 Dataset Description

| Dataset | Transactions | Classes | Items |
|---------|--------------|---------|-------|
| Breast | 699 | 2 | 18 |
| Car | 1728 | 4 | 25 |
| Ecoli | 336 | 8 | 34 |
| Glass | 214 | 7 | 20 |
| Iris | 150 | 3 | 12 |
| Nursery | 12960 | 5 | 27 |
| Pima | 768 | 2 | 36 |
| Zoo | 101 | 7 | 34 |

TABLE 4 Accuracy Comparison

| DATASET | IGmax Attribute | CBA (conf-50 sup-1) | GARC (Sup >1%, Conf > 50) | CWAC |
|---|---|---|---|---|
| Breast | 2 | 92.84 | **98.54** | 92.67 |
| Car | 6 | **90.51** | 88.50 | 78.33 |
| Ecoli | 6 | 80.36 | 69.27 | **86.14** |
| Glass | 8 | 61.68 | 61.90 | **76.19** |
| Iris | 4 | 93.33 | 94.60 | 94.60 |
| Nursery | 8 | 80.10 | 83.76 | **91.44** |
| Pima | 2 | 73.18 | 76.63 | **82.50** |
| Zoo | 3 | 83.18 | 85.14 | **88.00** |
| **Average** | | **81.90** | **82.29** | **86.23** |

TABLE 5 Number of rules generated

| DATASET | CBA (Sup>1%, conf >50%) | GARC (Sup >1% Conf>50%) | CWAC |
|---|---|---|---|

| Breast | 1912 | 447 | 312 |
|---|---|---|---|
| Car | 993 | 313 | 298 |
| Ecoli | 1116 | 375 | 311 |
| Glass | 4859 | 906 | 850 |
| Iris | 194 | 109 | 74 |
| Nursery | 3897 | 3788 | 1095 |
| Pima | 1520 | 356 | 251 |
| Zoo | 75611 | 4367 | 2695 |
| **Average** | **11262.75** | **1332.63** | **735.75** |

TABLE 6 Accuracy Comparison

| DATASET | GARC (Sup > 1%, Conf >70) | CWAC |
|---|---|---|
| Breast | 88.82 | **98.54** |
| Car | 81.57 | 79.02 |
| Ecoli | 70.48 | **87.95** |
| Glass | 77.14 | 76.19 |
| Iris | 94.60 | **96.00** |
| Nursery | 88.90 | **91.44** |
| Pima | 80.15 | 81.20 |
| Zoo | 86.00 | 86.00 |
| **Average** | **83.46** | **87.04** |

TABLE 7 Number of rules generated

| DATASET | GARC (Sup >1%, Conf > 70) | CWAC |
|---|---|---|
|  |  |  |

| Breast | 424 | 261 |
| --- | --- | --- |
| Car | 365 | 184 |
| Ecoli | 250 | 224 |
| Glass | 643 | 593 |
| Iris | 110 | 73 |
| Nursery | 1197 | 762 |
| Pima | 285 | 138 |
| Zoo | 3897 | 678 |
| **Average** | **896.38** | **364.13** |

Table 4 shows the accuracy comparison for various dataset with minimum support of 1%, minimum confidence is greater than 50%. The proposed algorithm has about +4.33 percent improvements against the traditional associative classification and about +3.94 percent improvements against the GARC respectively.

Table 5 shows number of rules generated. The traditional system has generated 15.31 times more rules than the proposed system and the GARC algorithm generated about 1.81 times more rules than the proposed system. If we consider the zoo dataset, it consists of only 101 transactions but it has 18 attributes and 7 class attribute, this leads to generation of more number of rules. On the other hand, the proposed system eliminates the un-necessary rules in the generation phase using gain attribute as well as link based weight, so it generates only minimal number of rules.

For each dataset, the parameter of the algorithm such as minimum support and minimum confidence varies to yield the best classifier. Obviously, the best setting for one dataset is different from the other. A recent work [7, 25] suggests the best minimum support and minimum confidence values as 1% and 70% respectively. Table 6 shows the accuracy comparison with minimum support of 1%, minimum confidence is equal to 70%. The proposed compact weighted associative classification method has about +3.58 percent improvements against the information gain based class association rule mining. Table 7 shows number of rules generated for minimum support 1%, minimum confidence of 70% and minimum chi square greater than of 3.84. The traditional system has generated 2.46 times more rules than the proposed system.

The proposed system generates lesser number of rules, on the same time it improves the accuracy of the system. So it is very easy to construct the classifier and to predict the new labels.

**Conclusion**

This paper proposes compact weighted class association rule mining method. The CWAC algorithm aims to extract the weighted class association rules from the dataset. Weight is computed using HITS algorithm, which does not require any preassigned weights. The proposed CWAC algorithm chooses information gain attribute and generates all the rules based on that attribute. So it generates compact ruleset. Experiment results shows that the proposed system not only generated lesser number of rules but also increases the classification accuracy. The process

International Journal of Data Mining & Knowledge Management Process (IJDKP) Vol.1, No.6, November 2011

of rule generation and rule evaluation can be further enhanced by implementing other attribute selection measures and rule evaluation strategies.

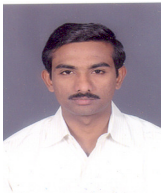
S. P. Syed Ibrahim received his BE in 2004 from Bharathidasan university and ME in 2006 from Anna University, Chennai. He has been research scholar of Anna Univerisy – Coimbatore. He is currently as Assistant professor in the department of Computer Science and Engineering at PSG College of Technology, Coimbatore. He has published more than 15 research papers in various journals and conferences. His research interest includes data mining, storage management.

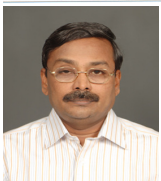
K.R. Chandran received his BE in 1982 and ME in 1986 from PSG College of Technology and PhD in 2006 from Bharathiar university, Coimbatore. He is currently professor in the department of Information Technology at PSG College of Technology, Coimbatore. He is a Fellow member in Institution of Engineers (India) and life member of ISTE, ACS, and The Textile Association (INDIA). He has published more than 55research papers in various journals and conferences. His research interest includes software engineering, computer networks, data mining, system modeling and simulation.